\documentclass[a4paper,fleqn,usenatbib]{mnras}

\usepackage{newtxtext,newtxmath}
\usepackage[T1]{fontenc}
\usepackage{ae,aecompl}

\usepackage{graphicx}	% Including figure files
\usepackage{amsmath}	% Advanced maths commands
\usepackage{amssymb}	% Extra maths symbols

\title[Dispersion inside active regions]{Dispersion of small magnetic elements inside active regions on the Sun}

\author[V.I.Abramenko]{
Valentina I. Abramenko\thanks{E-mail: vabramenko@gmail.com (VIA)}
\\
Crimean Astrophysical Observatory, Russian Academy of Science, Nauchny, Bakhchisaray,  298409, Crimea, Russia
}

\date{Accepted XXX. Received YYY; in original form ZZZ}

\pubyear{2015}

\begin{document}
\label{firstpage}
\pagerange{\pageref{firstpage}--\pageref{lastpage}}
\maketitle

\begin{abstract}

A process of diffusion of small-scale magnetic elements inside four active regions (ARs) was analyzed.    Line-of-sight magnetograms acquired by the Helioseismic and Magnetic Imager (HMI) onboard the Solar Dynamic Observatory (SDO) during a two-day time interval around the AR culmination time were utilized. Small magnetic elements of size of 3-100 squared HMI pixels with the field strength above the detection threshold of 30 Mx sm$^{-2}$ were detected and tracked. The turbulent diffusion coefficient was retrieved using the pair-separation technique. Comparison with the previously reported quiet-sun (QS) diffusivity was performed. It was found that: i) dispersion of small-scale magnetic elements inside the AR area occurs in the regime close to normal diffusion, whereas well-pronounced super-diffusion is observed in QS; ii) the diffusivity regime operating in an AR (the magnitude of the spectral index and the range of the diffusion coefficient) does not seem to depend on the individual properties of an AR, such as total unsigned magnetic flux, state of evolution, and flaring activity. We conclude that small-scale magnetic elements inside an AR do not represent an undisturbed photosphere, but they rather are  intrinsic part of  the whole coherent magnetic structure forming an  active region.  Moreover, turbulence of small-scale elements in an AR is not closely related to processes above the photosphere, but it rather carries the footprint  of the sub-photospheric dynamics.

\end{abstract}


%%%%%%%%%%%%%%%%%%%%%%%%%%%%%%%%%%%%%%%%%%%%%%%%%%

%%%%%%%%%%%%%%%%% BODY OF PAPER %%%%%%%%%%%%%%%%%%

\section{Introduction}

It is widely accepted that the solar magnetic field is generated by a dynamo when the field is maintained by plasma flows within the Sun \citep[][]{Moffatt1978, Parker1979}. Turbulent nature of the plasma flows involves a treatment of the magnetic field dispersion in the framework of turbulent diffusivity  \citep[][]{Parker1979}. Turbulent diffusivity is a key ingredient of any dynamo theory as a parameter that determines an interplay between the plasma flows and the magnetic field evolutionary changes on all temporal and spatial scales.
Thus, \citet[][]{Kitchatinov2013}  analyzed  differential rotation and meridional flows and  concluded that the dynamo number is in inverse proportion to the turbulent diffusivity, and that high temperature stars  should have higher values of turbulent diffusion coefficient.  
Nelson and co-authors \citep[][]{Nelson2013}, based  on  3D magnetohydrodynamic (MHD) simulations of convection and dynamo action in solar-type stars, concluded that  reducing of the turbulent diffusivity coefficient in the models leads to appearance of a cyclic reversal and an increase of the Reynolds number. The latter implies an enhanced level of turbulence and an increase of intermittency  in the global toroidal magnetic wreaths in both hemispheres of a star, which facilitates formation of buoyant loops at various depths in the convective zone, i.e., the turbulent-induced magnetic buoyancy.

Variations of the turbulent diffusion coefficient (frequently noted as $\eta_t$ ) with depth is, to some extent, a free parameter of a model. For example, in  flux transport dynamo models \citep[][]{Dikpati_Char1999, Dikpati_Gilm2006} the coefficient is adopted to be decreasing with depth from approximately 200 km$^2$ s$^{-1}$ on the surface and down to approximately 5  km$^2$ s$^{-1}$ in the bulk of the CZ below 0.9 $R_{\odot}$. To the contrary, in the mean field solar dynamo model with a double-cell meridional circulation \citep[][]{Pipin2011} the magnitude of  $\eta_t$ first increases from the surface to the depth of about 0.85 $R_{\odot}$, while  the decrease is gradually set on depths below 0.8 $R_{\odot}$.

In the majority of the existing dynamo models, the turbulent diffusivity coefficient $\eta_t$ is treated as a constant parameter over the solar surface and over all spatial and temporal scales. An exception  is the  theoretical and numerical approach suggested by Brandenburg and colleagues \citep[][]{Brand2008}. They defined a more general scale-dependent version of the transport coefficient: the turbulent magnetic diffusivity for the mean magnetic fields decreases monotonously with the increasing  wavenumber (in other words, with decreasing  spatial scales). Such variations of  $\eta_t$ with scales imply the turbulent regime of super-diffusivity, which was discovered recently in observations of the quiet sun magnetic fields \citep[][]{Abramenko2011, Lepreti2012, Gianna2013, Gianna2014a, Gianna2014b, Keys2014,  Yang2015, Jafar2017, Abramenko2017}.
 
Moreover, observed data showed that the regime itself varies with scales. In particular,  super-diffusivity is very strong and well pronounced on smallest observable scales ($\gamma = 1.5$ on scales 30-400 km),  it  is very low and growing from $\sim$ 20 to $\sim$150 $km^2 s^{-1}$ as spatial scales increase \citep[][]{Abramenko2011, Lepreti2012}, and  it becomes moderate on scales of 500-6000 km with $\gamma \approx 1.3$ and the magnitude  of the turbulent diffusion coefficient of 100-300 $km^2 s^{-1}$ \citep[][]{Abramenko2017}. It was also shown that the super-diffusion regimes operating in both  weakest  (a coronal hole, CH) and in moderate quiet-sun (super-granulation, SG) magnetic environments are quite similar \citep[][]{Abramenko2017}. This allows us to suggest that a unique mechanism generates and transports  the small-scale magnetic field inside vast solar areas outside active regions (ARs).  

Nevertheless, still remains open the question as to how does the dispersion of small magnetic elements inside ARs occur? In case these small elements are a part of the undisturbed photosphere, they should display the quiet-sun super-diffusion characteristics. If, however, the small-scale fields in an AR including the granulation field outside plage regions are part of the AR magnetic environment,  then the diffusivity characteristics may differ. If the diffusivity characteristics are indeed different,  then properties of small-scale field dispersion and transport may be influenced and carry some footprint of  sub-photospheric turbulent diffusion existing at depths from where the AR magnetic fields began to rise.  

The aim of the present study is to  measure magnetic  diffusivity of small magnetic elements inside  ARs and to compare the results with those  obtained previously for the undisturbed photosphere \citep[][]{Gianna2014a,Gianna2014b,Abramenko2017}.

\section{Data and method}

We utilized line-of-sight magnetograms (hmi.M-720s series)  acquired by the Helioseismic and Magnetic Imager (HMI) onboard  Solar Dynamic Observatory (SDO). The magnetograms were taken in the FeI 6173.3~A spectral line with the spatial resolution of 1 arcsec (pixel size of 0.5 arcsec),  cadence of 12~min \citep{Schou2012} and noise level of about 6  Mx sm$^{-2}$ \citep{Liu2012}.

\begin{figure*}
 	\includegraphics[width=\columnwidth]{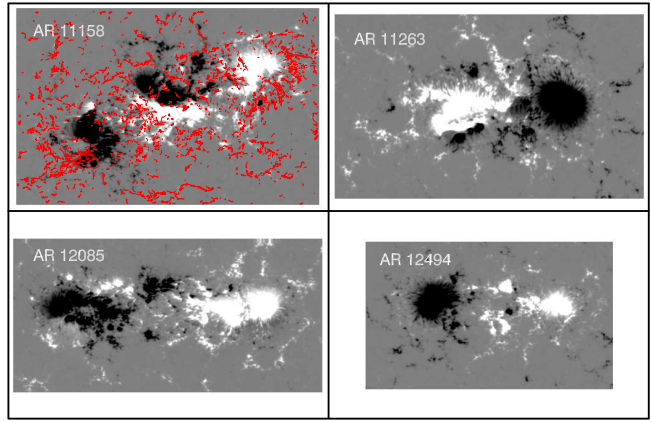}
 	\caption{\sf SDO/HMI magnetograms for four selected ARs. The magnetograms are scaled from -500 Mx c m$^-2$  (black) to 500 Mx c m$^-2$ (white). The width of the AR 12085-magnetogram is 500 pixels, or 180 Mm on the solar surface. All ARs are presented in the same scale at the time closest to their culmination (second column in Table 
 	\ref{tab1}).  
  	 For AR 11158, trajectories of all detected  magnetic elements are overplotted in red. Note, that the active region emerged during the observation period, so that trajectories of some elements became further overlapped by sunspots.}   
 	\label{fig-1}  
 \end{figure*}

Four ARs were selected for the analysis, Figure \ref{fig-1}. Two of them (NOAA AR 11158 and  12085) were observed  during their emergence and the other two ARs were rather stable. Similarly to the technique used in \citet[][]{Abramenko2017}, for each AR we prepared a two-day data set of 240 magnetograms each, centered on the moment of the AR culmination. The AR details  as well the calculated parameters, are listed in Table \ref{tab1}. Flare index, FI, of an AR (column 6) was calculated using  GOES data and  the approach suggested in \citet[][]{Abr2005}, where FI=1 (100) corresponds to a specific flare productivity of one C1.0 (X1.0) flare per day (FI data  starting from 2003 can be found at  http://solar.dev.argh.team/sunspots). In the same column 6, we also indicate the strongest flare that occured in an AR during its passage across the solar disk.

 %###########################################################
 %#############################################################

\begin{table*}
	\caption{Calculated parameters}
	\label{tab1}
	\begin{tabular}{lccccccccc}
		\hline
		NOAA &    Culmination &  Lat, & Flux$^a$,& Stage  & Flare Index/  &   $K$,         & $\gamma$ & $N^b$   \\  
	  	     &                &  deg  &$10^22$ Mx&        &Strongest flare& km$^2$ s$^{-1}$&          &         \\ 
		\hline
		11158 &2011.02.14/00UT& S19   & 1.1-2.7  &Emerging&  41.5/X2.2    &  120 - 200  & 1.168$\pm$0.012& 1616  \\
		11263 &2011.08.03/12UT& N17   & 2.2-2.2  & Stable &  55.0/X6.9    &  113 - 180  & 1.157$\pm$0.014& 1447  \\
		12085 &2014.06.09/09UT& S20   & 1.2-2.2  &Emerging&  11.0/M1.8    &  122 - 130  & 1.022$\pm$0.015&  953  \\
		12494 &2016.02.05/18UT& S12   & 1.0-1.0  & Stable &  1.0/C3       &  145 - 242  & 1.171$\pm$0.017& 1263  \\
		
		\hline
		\multicolumn{4}{l}{$^a$ range of the total unsigned flux changes}\\
		\multicolumn{9}{l}{$^b$ Number of tracked elements}\\
	\end{tabular}
\end{table*}

Since we intend  to compare the diffusion regime in ARs with that  in the undisturbed photosphere, we  follow the same detection and tracking routine that was applied to  the quiet-sun (QS) areas and described in details in \citet[][]{Abramenko2011,Lepreti2012,Abramenko2017}.The thresholding technique was applied there to the modulus of the magnetic field strength to select  magnetic flux concentrations with  size from 3 to 100 squared pixels.  In  QS-areas,  threshold  of 20 Mx sm$^{-2}$ was used  to outline a sufficiently large number of elements with field strengths that   exceeded triple standard deviation (6 Mx sm$^{-2}$). However, when considering AR  areas, we found only few cases in  a magnetogram that satisfy the above requirements. To detect a sufficient number of small-scale elements (at least  100 per magnetogram)  of size 3-100 squared pixels, we were forced to increase the threshold level to 30 Mx sm$^{-2}$. This experiment indicates that the overall magnetic flux density  on small scales inside ARs is higher than that in QS-areas. This inference  will be further  confirmed below by an analysis of magnetic power spectra.

\begin{figure*}
 	\includegraphics[width=\columnwidth]{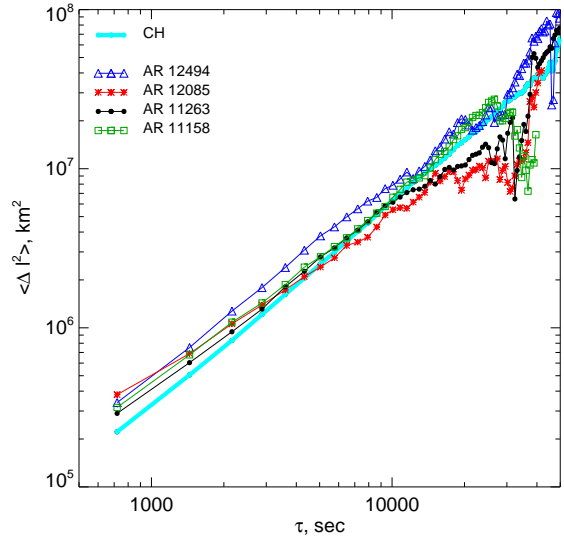}
 	\caption{\sf The AR displacement spectra $ \langle(\Delta l)^{2}\rangle $ acquired via the pair-separation technique. The linear range, where the spectral index $\gamma$ was calculated, covers the time interval of 720-14400 s, or 12 min - 4 hours. For comparison, the separation spectrum for the coronal hole area recorded on 2016 January 2-4 \citep[][]{Abramenko2017} is shown with a thick turquoise line.  }
 	\label{fig-2}  
 \end{figure*}

About one thousand  magnetic elements for each AR were detected and tracked (see  last column in  Table \ref{tab1}). Note that elongated chains of several local maxima above the threshold level were not considered.  By tracking the selected elements we first generated their trajectories that were then used to derive the displacement spectra. The spectra were calculated using the pair-separation technique  \citep[][]{Monin1975,Lepreti2012}, when displacements are computed as an increment in the distance between two tracers at consecutive moments (see Eq.2 in \citet[][]{Abramenko2017}). As compared to the traditional single-particle displacement spectrum, the two-particle dispersion reflects diffusivity properties arising from the inertial range of turbulence, thus allowing  us to avoid possible influence of large-scale velocity patterns.   The displacement spectrum, $\langle(\Delta l)^{2}(\tau_i)\rangle$, as a function of the time interval $\tau$ is  shown in Figure \ref{fig-2} for all ARs. The sample interval, $\tau_i$, is measured from the first appearance of both tracers in the pair, and the displacements for a given $\tau_i$ are then averaged over all pairs.

 %###########################################################
 %#############################################################

A linear range between 720 s and 14400 s (12 min - 4 h) is well pronounced in all spectra shown in Figure \ref{fig-2}. Inside this interval we calculated the best linear fit, and the corresponding  slope of the fit or spectral index, $\gamma$, is shown in the 8th column of Table \ref{tab1}. The spectral index and the y-intercept of the fit allowed us to compute the turbulent diffusion coefficients, $K(\tau)$, and $K(\Delta l)$ by using Eqs. 6-7 in \citet[][]{Abramenko2017}. (Hereinafter, the turbulent diffusion coefficients is denoted as $K$.)

\section{Results}

\begin{figure*}
 	\includegraphics[width=\columnwidth]{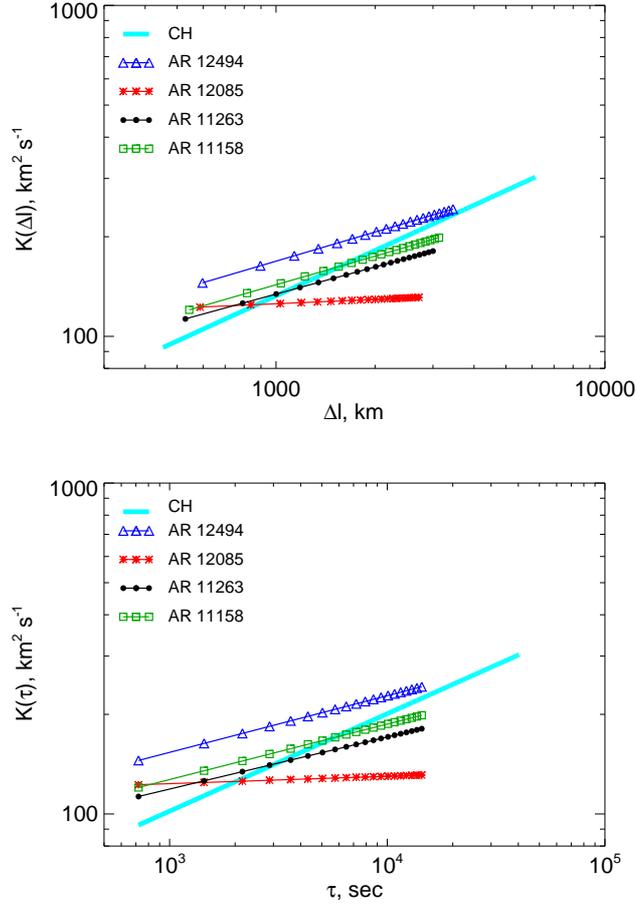}
 	\caption{\sf Turbulent diffusivity coefficients for four ARs plotted  as a function of temporal (bottom) and spatial (top) scales. For comparison, the turbulent diffusivity coefficient calculated for a coronal hole (CH)  in \citep{Abramenko2017} is shown with a thick turquoise line.    }
 	\label{fig-3}  
 \end{figure*}

 The calculated turbulent diffusion coefficients are shown in Figure \ref{fig-3} as a function of temporal and spatial scales. In both graphs the $K$ plots  are stacked on top of each other along the vertical axis, however, no correlation with the general AR parameters is observed. Indeed, the spectra of the two emerging ARs (NOAA 11158 and 12085) do not appear distinct and they are similar to  the spectra of the stable ARs (NOAA 11263 and 12494), while  the spectra of two the most flare-productive ARs (NOAA 11156 and 11263) are nested in between the spectra of the low-flaring ARs (NOAA 12085 and 12494) and show no specific features.  

 %###########################################################
 %#############################################################
 %###########################################################
 \begin{figure*}
 	\includegraphics[width=\columnwidth]{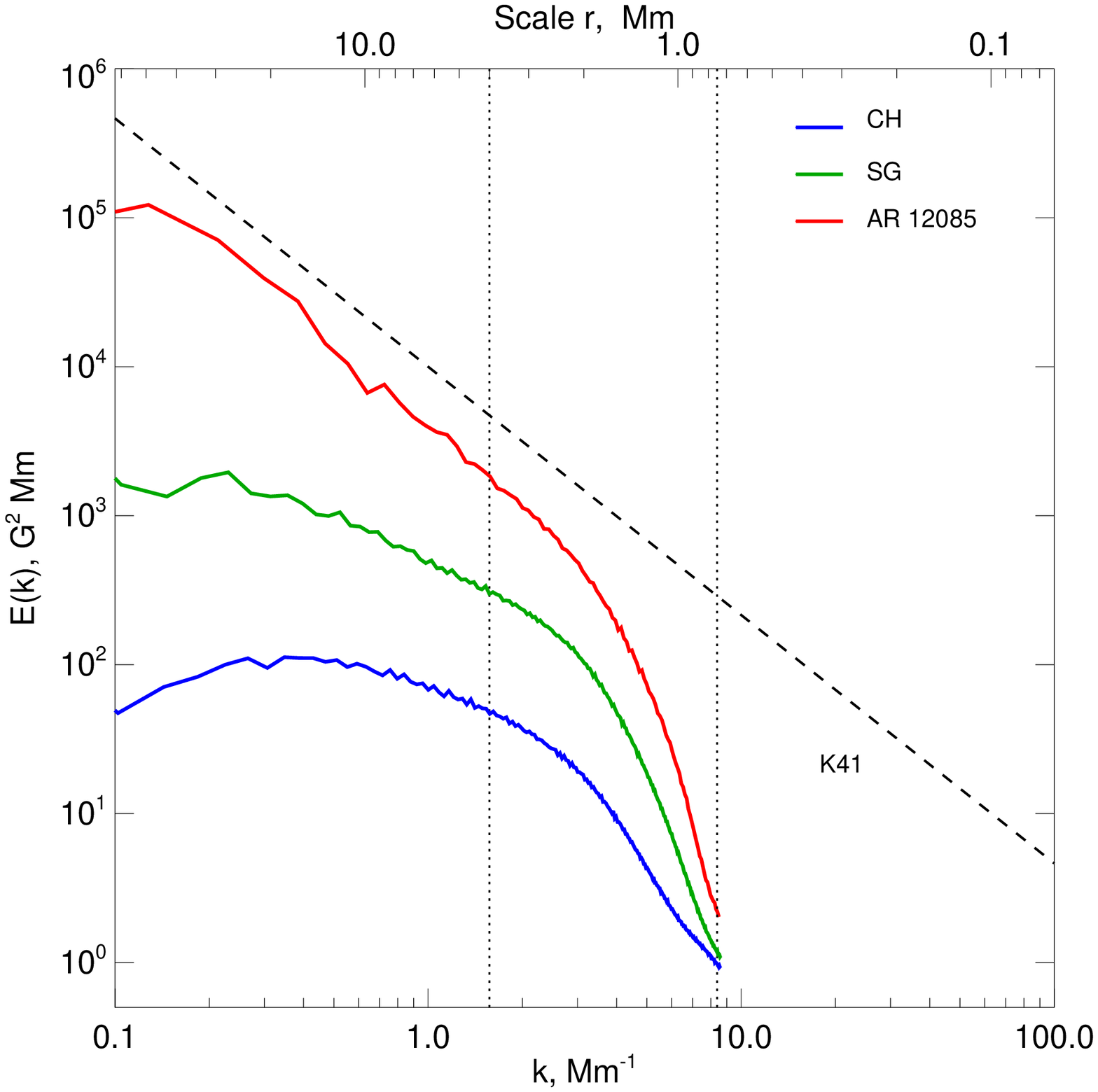}
 	\caption{\sf Averaged over 240 magnetograms magnetic power spectra for  NOAA AR 12085 (red),  a coronal hole area (blue), and  a super-granulation area (green). The quiet Sun data are described in \citet[][]{Abramenko2017}. The dashed  K41 line represents  Kolmogorov -5/3 spectrum. The scale interval of the present analysis is indicated  by the vertical dotted lines.  }
 	\label{fig-4}  
 \end{figure*}
 %#############################################################

At the same time, there exists a well-pronounced difference between the quiet  Sun and active regions diffusivity.  Data for a coronal hole  were adopted from \citet{Abramenko2017} and overplotted in Figure \ref{fig-3}  for comparison with a thick turquoise line. We thus note weaker super-diffusivity in ARs.  The increase of $K$ with scales is slower in ARs than in the QS-area. The spectral index $\gamma$ varies from 1.02 to 1.17 in  the ARs, whereas in the QS-areas $\gamma =1.3$ was reported by \citet{Abramenko2017}.  (Note, that  super-granulation  exhibits the same turbulent diffusion regime as the CH \citep{Abramenko2017}). Therefore, the dispersion of small-scale magnetic elements in AR, being closer to the normal diffusions,  notably differs from that in the quiet Sun.

The high degree of super-diffusivity implies presence of  high  degree of intermittency with frequent large fluctuations (Levy flights) and an efficient search strategy of chaotic displacements \citep{Klafter2005}, whereas  normal diffusion implies a non-correlated random walk with very rare large fluctuations. Small elements inside an AR are a subject of continuous strong buffeting by neighbouring large magnetic elements and flows, and this might be the reason for non-correlated displacements. This specific (different from QS) turbulent regime of small-scale elements in ARs suggests that they are an intrinsic part of the entire magnetic structure of an AR and, as such, they might belong to a turbulent cascade of an AR.
To explore this suggestion, we calculated  magnetic power spectra using AR magnetograms and QS-area data analysed in \citet{Abramenko2017}. The power spectra were computed using the method presented in \citet{Abr2005}.

A typical AR magnetic power spectrum overplotted with power spectra of  CH and SG areas  is shown in Figure \ref{fig-4}. Note, that according to \citet{Abr2005}, the cutoff of the magnetic power spectrum as determined from SOHO/MDI data occurs at scales of about 3 Mm, while   this plot shows that in the case of SDO/HMI data  the cutoff scale is at about 2.5 Mm. This HMI cutoff scale is now inside the interval of our present analysis (vertical dotted lines in Figure \ref{fig-4}). The instrumentation cutoff distorts the spectrum. Nevertheless, one can conclude that magnetic energy on small scales is systematically higher in AR than in QS, and small-scale elements in QS and in ARs belong to different turbulent cascades.

\section{Concluding remarks}

While super-diffusivity in an undisturbed photosphere is now  a well established  fact, the  nature of turbulent diffusion in ARs is still  an open question. Here we explored magnetic flux dispersion by using SDO/HMI magnetograms for four ARs obtained at different stage of evolution and different flare productivity, and compared the results with the dispersion properties measured for quiet Sun areas. We report the following.  

-  The dispersion process of small-scale magnetic elements inside AR differs from that of  the undisturbed photosphere. The regime of turbulent diffusion in ARs on scales of 500-4000 km is close to the normal diffusion whereas in the quiet Sun super-diffusivity with $\gamma$ = (1.3 - 1.4) is well pronounced \citep{Gianna2014a, Abramenko2017}. We found the magnitude of $\gamma$ varying in ARs in the range of 1.02-1.17. The turbulent diffusion coefficient, $K$, was found to be  120-130 km$^2$ s$^{-1}$ in the most ``normal-diffusion`` AR 12085 ($\gamma$ = 1.02), and it is in the range of 145-240 km$^2$ s$^{-1}$ for the most ``super-diffusive`` AR 12494 ($\gamma$ = 1.17).

- The diffusivity regime  as determined from the magnitude of $\gamma$ and the range of $K$, does not appear to  depend on individual properties of an AR, such as total unsigned magnetic flux, state of evolution, and flaring activity.

- Analyzed magnetic power spectra show that small-scale element in an AR may belong to the classical Kolmogorov's turbulent cascade with the spectral index close to -5/3. At the same time,  magnetic power spectra for QS areas are shallower with the spectral index about -1. The less steep spectra of CH and SG indicate that in quiet sun the ratio between small scale and large scale energy is higher with respect to what is found in ARs.	 

These conclusions  allow us to further suggest that:

i) Small-scale magnetic elements inside the AR area are not representative of the undisturbed photosphere, but they rather intrinsically belong to the whole coherent magnetic structure of an AR and probably are a result of the evolutionary processes in an AR such as flux emergence and dispersion .

ii) Turbulent properties of small-scale elements in an AR  are not defined by  processes above the photosphere, but they rather carry the footprint of the sub-photospheric dynamics. In this case analyzing  turbulence inside ARs, we have a chance to probe  deep layers of the convection zone. As it follows from Figure \ref{fig-3}, the magnitude of $K$ at smallest observable scales is higher in ARs than in QS-areas and in ARs it decreases slower with decreasing spatial scale as compared to QS-areas. The QS turbulence is a near-surface phenomenon. If the AR turbulence regime indeed carries a footprint of the deep-layer dynamics, then  we may further speculate  that the coefficient of turbulent diffusivity should increase with depth, which is in agreement with the modeling approach presented in \citet{Pipin2011}. Further analysis based on larger statistics might allow us to estimate the depth at which an ARs are rooted in the convection zone, which is an important ingredient of various dynamo models \citep[e.g.,][]{Nelson2013}.

\section*{Acknowledgements}

SDO is a mission for NASA Living With a Star (LWS) program. The SDO/HMI data were provided by the Joint Science Operation Center (JSOC). I am thankful to anonymous referee for useful comments helping to improve the paper. The study was supported in part by the Russian Foundation for Basic Research projects 16-02-00221 A and 17-02-00049. 

%%%%%%%%%%%%%%%%%%%%%%%%%%%%%%%%%%%%%%%%%%%%%%%%%%



%\bibliographystyle{mnras}


%%%%%%%%%%%%%%%%%%%%%%%%%%%%%%%%%%%%%%%%%%%%%%%%%%

% Don't change these lines
\bsp	% typesetting comment
\label{lastpage}
\end{document}